\documentclass[10pt,conference]{IEEEtran}
\usepackage{times}
\usepackage[cmex10]{amsmath}
\usepackage{amssymb}
\usepackage{epsfig,verbatim}
\usepackage{algorithm}

\usepackage{algpseudocode}
\long\def\comment#1{}

\newfont{\bbb}{msbm10 scaled 700}

\newfont{\bb}{msbm10 scaled 1100}
\newcommand{\CC}{\mbox{\bb C}}
\newcommand{\PP}{\mbox{\bb P}}
\newcommand{\RR}{\mbox{\bb R}}

\newcommand{\bv}{{\bf b}}
\newcommand{\cv}{{\bf c}}

\newcommand{\hv}{{\bf h}}

\newcommand{\qv}{{\bf q}}
\newcommand{\rv}{{\bf r}}

\newcommand{\wv}{{\bf w}}

\newcommand{\xv}{{\bf x}}
\newcommand{\yv}{{\bf y}}
\newcommand{\zv}{{\bf z}}

\newcommand{\Hm}{{\bf H}}

\newcommand{\Bc}{{\cal B}}
\newcommand{\Cc}{{\cal C}}

\newcommand{\Ec}{{\cal E}}

\newcommand{\Nc}{{\cal N}}

\newcommand{\Sc}{{\cal S}}

\newcommand{\Wc}{{\cal W}}

\newcommand{\alphav}{\hbox{\boldmath$\alpha$}}

\newcommand{\eqdef}{\stackrel{\Delta}{=}}

\newcommand{\transp}{{\sf T}}

\newtheorem{definition}{Definition}

\newtheorem{remark}{Remark}

\newcommand{\argmax}{\operatornamewithlimits{argmax}}

\title{Successive Cancellation Soft Output Detector For Uplink MU-MIMO Systems With One-bit ADCs}

\author{
\IEEEauthorblockN{
              Yun-Seong Cho, Seonho Kim and Song-Nam Hong}
\IEEEauthorblockA{Ajou University, Suwon, Korea,\\
              email: \{cys4577, kimsh1005, snhong\}@ajou.ac.kr}
              }

\begin{document}

\maketitle

%\date{}

%\blfootnote{
% and  US Patent No. 14/824694 "Rate-Compatible Polar Codes" was filed on Aug.~12, 2015.
%}

%%%%%%%%%%%%%%%%%%%%%%%%%%%%%%%%%%%%%%%%%%%%%%%%%%%%%%%%%%%%%%%%%%%%%%%%%%%%%%%%%%%%%%%%%
\begin{abstract} 
In this paper, we present a successive-cancellation-soft-output (SCSO) detector for an uplink multiuser multiple-input-multiple-output (MU-MIMO) system with one-bit analog-to-digital converters (ADCs). The proposed detector produces soft outputs (e.g., log-likelihood ratios (LLRs)) from one-bit quantized observations in a {\em successive} way: each user $k$'s message is sequentially decoded from a channel decoder $k$ for $k=1,...,K$ in that order, and the previously decoded messages are exploited to improve the reliabilities of LLRs. Furthermore, we develop an efficient greedy algorithm to optimize a decoding order. Via simulation results, we demonstrate that the proposed ordered SCSO detector outperforms the other detectors for the coded MU-MIMO systems with one-bit ADCs.
\end{abstract}

%%%%%%%%%%%%%%%%%%%%%%%%
\section{Introduction}
%%%%%%%%%%%%%%%%%%%%%%%%%%%%

A massive multiple-input-multiple-output (MIMO) is one of the promising techniques to cope with the predicted wireless data traffic explosion \cite{Marzetta}-\cite{Lu}. Whereas, the massive MIMO can considerably increase the hardware cost and  the radio-frequency (RF) circuit consumption \cite{Yang}. Among all the components in a RF chain, a high-resolution analog-to-digital converter (ADC) is particularly power-hungry as the power consumption of an ADC is scaled exponentially with the number of quantization bits and linearly with the baseband bandwidth \cite{Murmann} and \cite{Mezghani-2011}. To overcome this challenge, low-resolution ADCs  (e.g., 1$\sim$3 bits) for massive MIMO systems have been considered as a low-power solution over the past years. The one-bit ADC is particularly attractive due to its lower hardware complexity.

%%% Recent Works %%%

Numerous MIMO detectors have been developed for uplink MU-MIMO systems with one-bit ADCs. The optimal maximum-likelihood (ML) detector was introduced in \cite{Choi} and some low-complexity detectors were also proposed in \cite{Choi}-\cite{Mollen2}. Also, new MIMIO detection frameworks based on supervised learning and coding theory were recently presented in  \cite{Lee} and \cite{Hong_J}, respectively. In spite of their attractive {\em uncoded} performances, they yield a poor performance for the {\em coded} systems since their hard-decision outputs degrade the performances of the following channel codes  (e.g., Turbo \cite{Berrou}, low-density-parity-check (LDPC) \cite{Richardson} and polar codes \cite{Arikan}). 

%%%%%%%%%%%%%%
In our recent work, the above problem has been addressed in \cite{Hong-Soft} by presenting a soft-output (SO) detector. In this method, the soft values are derived from the hard-decision observations by exploiting a novel distance measure, named weighted Hamming distance. Therefore, the SO detector can be naturally incorporated into a state-of-the-art {\em soft} channel decoder (e.g., belief-propagation decoder \cite{Richardson}) while the hard-output detectors in \cite{Choi}-\cite{Hong_J}  should be combined with highly suboptimal {\em hard} channel decoder (e.g., bit-flipping decoder \cite{Rao}). In \cite{Hong-Soft}, it was shown that the SO detector provides a substantial coded gain (about 10dB) over the optimal (hard-output) ML detector.

%%%%%%%%%%%%%%%%%%%%%%%%%%%%%%%%%%%
\begin{figure}
\centerline{\includegraphics[width=9cm]{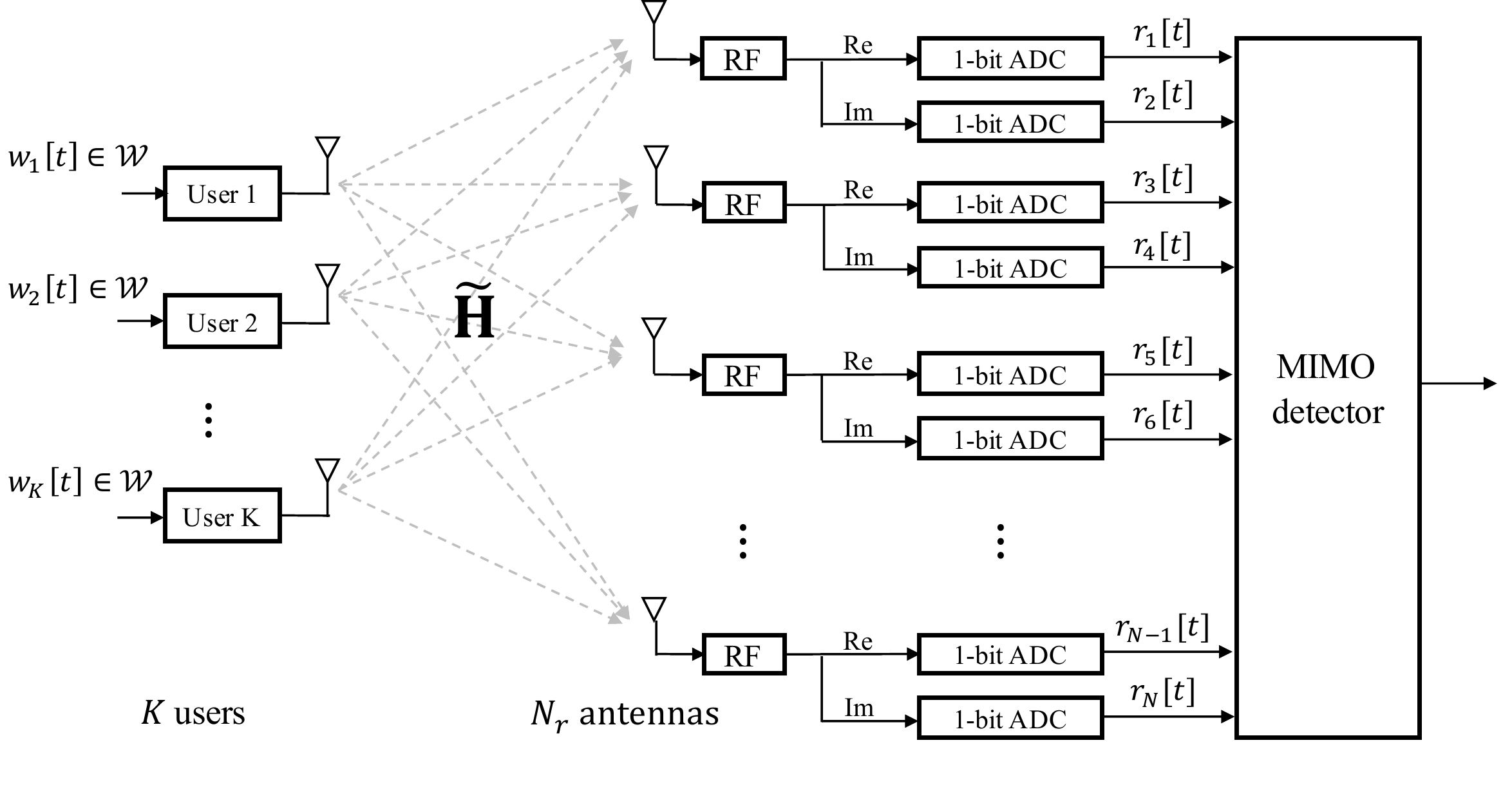}}\vspace{-0.5cm}
\caption{Illustration of an uplink multiuser massive MIMO system in which each receiver antenna at a BS is equipped with one-bit ADC.}
\label{systemmodel}
\end{figure}

%%% Our Contributions %%%

In this paper, we propose a {\em successive-cancellation-soft-output} (SCSO) detector which enhances the SO detector in \cite{Hong-Soft} by exploiting a {\em priori} knowledge. In the SO detector, the LLRs are computed using the relative distances among the current observation and all possible {\em noiseless} channel outputs (say, codewords) where the set of such codewords are referred to as a spatial-domain code $\Cc$. The key difference of the proposed SCSO detector is that the code $\Cc$ is refined using a priori knowledge. To be specific, the SCSO detector produces the LLRs in a {\em successive} way: each user $k$'s message is sequentially decoded from a channel decoder $k$ for $k=1,...,K$ in that order and the soft inputs (LLRs) of the channel decoder $k$ are computed from the {\em refined} code. Here, the refined code is constructed by eliminating some codewords from the $\Cc$ using the previously decoded messages. Since the codewords in the refined code can have larger distances than those in the $\Cc$, a detection ambiguity can be reduced. We further improve the proposed detector by optimizing a decoding order. We notice that due to the non-linearity of the effective channel, it is not possible to employ the ordering idea in V-BLAST \cite{Wolniansky}. Instead, we determine a decoding order in a greedy fashion such that the resulting subcodes have a better structure, i.e., the distances of the remaining codes are as far as possible. Via numerical results, it is demonstrated that the ordered SCSO detector can attain 1.5dB coded gain over the SO detector for the polar coded MU-MIMO systems with one-bit ADCs.

%%%%%%%%%%
\textbf{Notation:} Lower and upper boldface letters denote column vectors and matrices, respectively. Let $[a]\eqdef\{1,...,a\}$ for any positive integer $a$. Also, for any vector $\xv$, let $\xv_{a}^{b}=[x_a, x_{a+1},\ldots,x_{b-1}, x_b]$ denote the part of the vector $\xv$ for some positive integers $a$ and $b$ with $b>a$. For any $\ell \in \{0,1,\ldots,L-1\}$, we let $g(\ell)=[b_0,b_1,\ldots,b_{L-1}]^{\transp}$ denote the $m$-ary expansion of $\ell$ where 
$\ell=b_0m^0+\cdots+b_{L-1}m^{L-1}$. We also let $g^{-1}(\cdot)$ denote its inverse function. For a vector, $g(\cdot)$ is applied in an element-wise manner. 
%%%%%%%%%%%%%%%%%%%%%%%%%%
\section{Preliminaries}\label{sec:pre}

In this section, we describe a system model and define its equivalent channel to be used in the sequel.

%%%%%%%%%%%%%%%%%%%%%%%
\subsection{System Model}\label{subsec:sm}

We consider a single-cell uplink MU-MIMO system in which there are $K$ single-antenna users and one base station (BS) equipped with an array of $N_r > K$ antennas. We use $t$ to indicate a time-index label. Let $w_k[t] \in \Wc=\{0, \ldots, m-1\}$ represent the user $k$'s message at time slot $t$ for $k \in [K]$, each of which contains $\log{m}$ information bits. We also denote $m$-ary constellation set by $\Sc=\{s_0,\ldots,s_{m-1}\}$. For the ease of expression, it is assumed that $m=2^p$ for some positive $p$.  However, the proposed detectors in this paper can be immediately extended to an arbitrary $m$. Then the transmitted symbol of the user $k$ at time slot $t$, $\tilde {x}_k(w_k[t])$ can be obtained by a modulation function $f$ as
\begin{equation}
\tilde {x}_k(w_k[t])=f(w_k[t]) \in \Sc.
\end{equation}
When $K$ users transmit the symbols $\tilde {\xv}(\wv[t])=[\tilde {x}_1(w_1[t]),\ldots,\tilde {x}_K(w_K[t])]$, the discrete-time complex-valued baseband received signal at the BS is 
\begin{equation}
\tilde{\rv}[t] = \tilde{\Hm}\tilde {\xv}(\wv[t]) +\tilde{\zv}[t] \in \CC^{N_r},\label{eq:inout} 
\end{equation} where $\tilde{\Hm} \in \CC^{N_r \times K}$ is the channel matrix between the BS and the $K$ users, i.e., $\hv_{i}^{\transp} \in \CC^{1 \times K}$ is the channel vector between the $i$-th receiver antenna at the BS and the $K$ users. In addition, $\tilde{\zv[t]}=[\tilde{z}_0[t],\ldots,\tilde{z}_{N_r}[t]]^\transp \in \CC^{N_r}$ is the noise vector whose elements are distributed as circularly symmetric complex Gaussian random variables with zero-mean and unit-variance, i.e., $\tilde{z}_i[t]\sim\Cc\Nc(0,1)$. 

In the MU-MIMO system with one-bit ADCs, each receive antenna of the BS has RF chain which consists of two one-bit ADCs that separately applied to real and imaginary part (see Fig.~\ref{systemmodel}). Let $\mbox{sign}(\cdot): \RR \rightarrow \{0,1\}$ represent the one-bit ADC quantizer function with
\begin{equation}
\mbox{sign}(\tilde{r}[t])=
\begin{cases}
0 & \mbox{ if } \tilde{r}[t] \geq 0\\
1 & \mbox{ if } \tilde{r}[t]  < 0.
\end{cases}
\end{equation} For a vector, it is applied element-wise. After applying ADC quantizers, the BS observes the quantized received output vector as
\begin{align*}
\hat{\rv}_{\rm R}[t] &= \mbox{sign}(\mbox{Re}(\tilde{r}[t]))\in \{0,1\}^{N_r}\\
\hat{\rv}_{\rm I}[t] &= \mbox{sign}(\mbox{Im}(\tilde{r}[t])) \in \{0,1\}^{N_r}.
\end{align*}

In this paper, we only consider a real-valued channel for the ease of representation, and we can remodel the complex-valued input-output relationship in (\ref{eq:inout}) into the equivalent real-valued representation as

\begin{equation}
\rv[t] = \mbox{sign}(\Hm\xv(\wv[t]) +\zv[t]) \in \RR^{2N_r}, \label{eq:real}
\end{equation}
where 
\begin{equation}
\begin{cases}
\Hm=\left[ {\begin{array}{cc}
   \mbox{Re}(\tilde{\Hm}) & -\mbox{Im}(\tilde{\Hm}) \\      
   \mbox{Im}(\tilde{\Hm}) & \mbox{Re}(\tilde{\Hm})\\
 \end{array} } \right] \in \RR^{2N_r \times 2K }\\
\rv[t]=[\hat{\rv}_{\rm R}[t]^{\transp}, \hat{\rv}_{\rm I}[t]^{\transp}]^{\transp}\\
\xv(\wv[t])=[\mbox{Re}(\tilde{\xv}(\wv[t]))^{\transp}, \mbox{Im}(\tilde{\xv}(\wv[t]))^{\transp}]\\
\zv[t]=[\mbox{Re}(\tilde{\zv})^{\transp}, \mbox{Im}(\tilde{\zv})^{\transp}].\\
 
\end{cases}
\end{equation}
This real system representation will be used in the sequel.

 A block fading channel is assumed where a channel matrix $\Hm$ remains constant during $n$ time slots (i.e., coded block length) and changes independently across coded blocks. Also, it is assumed that 
the channel matrix $\Hm$ is perfectly known to a BS. It is remarkable that the proposed soft-output detector can be also performed with an estimated channel matrix $\hat{\Hm}$ by simply changing $\Hm$ into $\hat{\Hm}$ in the following sections.

%%%%%%%%%%%%%%%%%%%%%%%%%%%%%%
\begin{figure}
\centerline{\includegraphics[width=9cm]{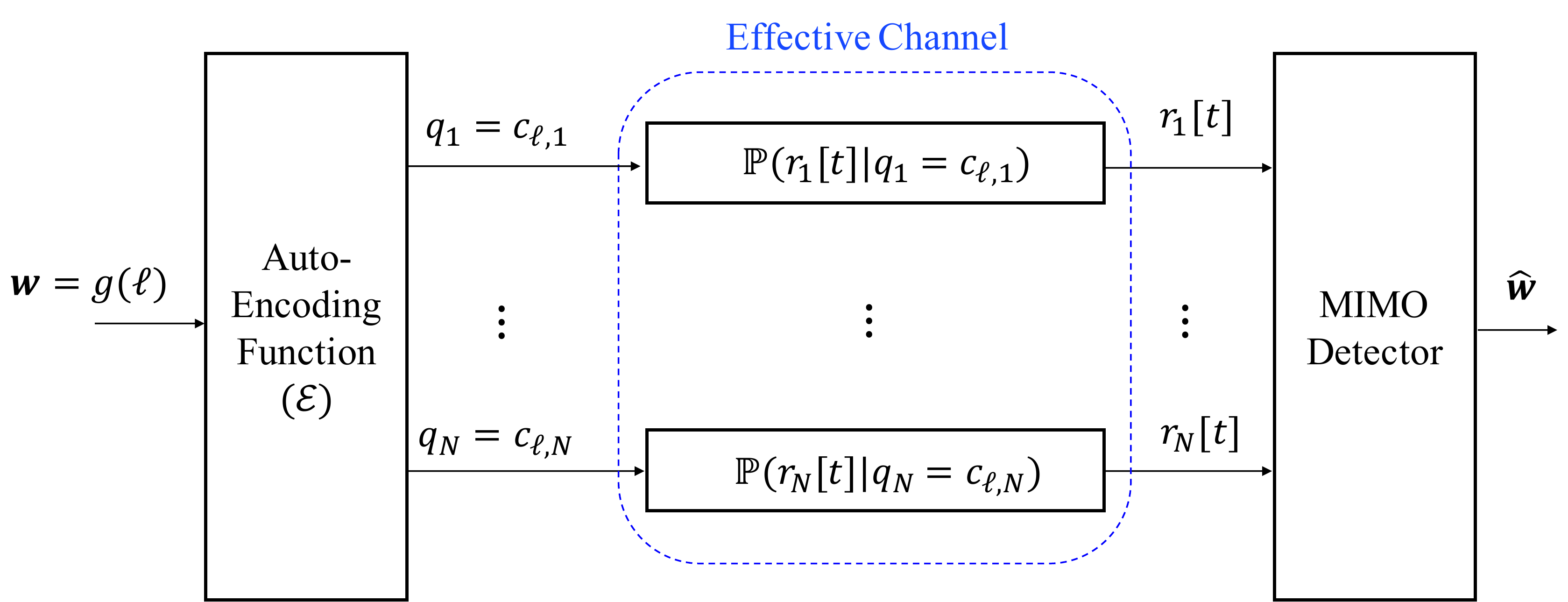}}
\caption{Equivalent $N$ parallel B-DMCs where an auto-encoding function $\Ec$ is determined as a function of $\Hm$ and the transition probabilities of the effective channel depend on the message vector $\wv$.}
\label{e_model}
\end{figure}

\subsection{Equivalent $N$ parallel B-DMCs} \label{subsec:EDMC}
In \cite{Hong_J}, it was shown that a real system representation in (\ref{eq:real}) can be transformed into an equivalent $N=2N_r$ parallel binary discrete memoryless channels (B-DMCs). In this section, we define the channel input/output and channel transition probabilities of the $N$ parallel B-DMCs (see Fig.~\ref{e_model}). Due to the equivalence, the channel output is clearly $\rv[t]$.

 %%%%%%%%%%%%%%%%%%%%%%%%%%%%%%%%%%%
\begin{figure*}
\centerline{\includegraphics[width=18cm]{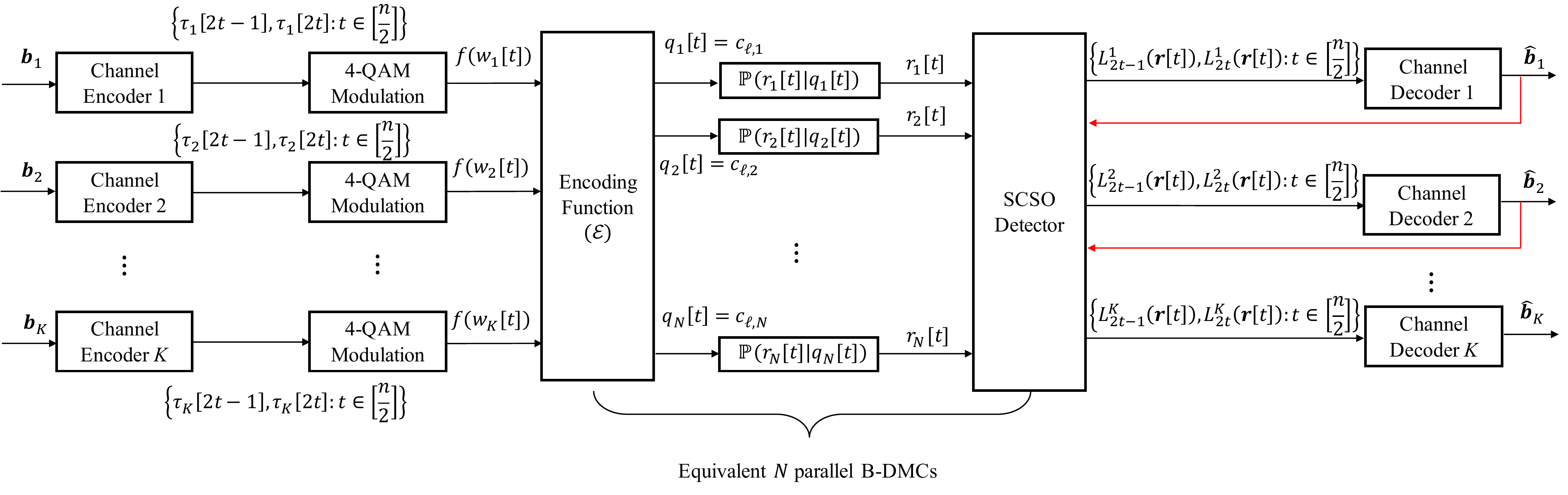}}
\caption{The proposed SCSO detector.}
\label{codedmodel}
\end{figure*}
%%%%%%%%%%%%%%%%%%%%%%%%%

{\em Channel input:} For a given $\Hm$, we define a spatial-domain code $\Cc=\{\cv_0,\ldots,\cv_{m^K-1}\}$
where each codeword $\cv_{\ell}$ is determined as a function of $\Hm$ as
\begin{equation}
\cv_{\ell}=[\mbox{sign}(\hv_1^\transp\xv(g(l)), \ldots, \hv_{N}^\transp\xv(g(l))]^\transp \in \Cc,\label{eq:code} 
\end{equation}
This code is solely based on \Hm. From Fig.~\ref{e_model}, the output of auto-encoding function $\Ec$ is generated by 
\begin{equation}
\qv=\Ec(\wv)=\cv_{\ell},\label{eq.enc}
\end{equation}
 where $\ell=g^{-1}(\wv) \in \{0,\ldots,m-1\}$\\

{\em Transition probability:} As shown in Fig.~\ref{e_model}, the effiective channel is composed of $N$ unequal parallel BSCs with input $\qv$ and output $\rv$. Also, for the $i$-th subchannel, the transition probabilities, depending on user's message $\wv=g(l)$ and  corresponding codeword $\cv_{\ell}$, are defined as
\begin{equation}
\PP(r_i[t] \mid q_i=c_{\ell,i})=\begin{cases}
\epsilon_{\ell,i} & \mbox{ if } r_i[t] \neq c_{\ell,i}\\
1-\epsilon_{\ell,i} & \mbox{ if }  r_i[t] = c_{\ell,i}, \label{eq:probdef}
\end{cases}
\end{equation}
where the error probability of the $i$-th channel is computed as
\begin{equation}
\epsilon_{\ell,i}= Q(|\hv_i^\transp\xv(g(\ell))|<0) \label{eq:cross_p}
\end{equation}
 and 
\begin{equation} 
 Q(x) = \frac{1}{2\pi}\int_{x}^{\infty} \exp\left(-u^2/2\right) du.
 \end{equation}

%%%%%%%%%%%%%%%%%%%%%%%%%%%%
\section{The SO Detector}\label{sec:swmd}

In this section, we review the soft-output (SO) detector proposed in \cite{Hong-Soft}. Some useful definitions that will be used throughout the paper are provided as follows.

%%%%%%%%%%%%%%%%%%%%%%%%%%%%
\begin{definition}[distance measure]\label{def:w_MD}
For any two vectors $\xv$ and $\yv$ of length $N$, we define a {\em weighted} Hamming distance $d_{wh}(\xv,\yv;\alphav)$  with the weights $\{\alpha_i\}_{i=1}^{N}$ as
\begin{equation}
d_{wh}(\xv,\yv; \{\alpha_i\}) \triangleq \sum_{i=1}^{N} \alpha_i \mathbf{1}_{\{x_i \neq y_i\}}.
\end{equation}
\end{definition}
%%%%%%%%%%%%%%%%%%%%%%%%%
\begin{definition}[subcode] \label{def:subcode} 
For any fixed $\{w_{k} = j\}$, the subcode of $\Cc=\{\cv= \Ec(\wv): \wv \in \Wc^{K}\}$ is defined as
\begin{align*}
\Cc_{|\{w_{k} = j\}} \eqdef \{\cv= \Ec(\wv): \wv \in \Wc^{K}, w_{k} = j\}.
\end{align*}
\end{definition}
\vspace{0.1cm}

We are now ready to explain how the SO detector computes the LLR values from the one-bit quantized observation $\rv[t]$. Fig.~\ref{codedmodel} describes the SO detector for $p=2$ (i.e., 4-QAM) where the red lines (i.e., feedbacks from channel decoders) are not used. Let $(\tau_k[1],...\tau_k[n])$ denote the coded output of the user $k$'s channel encoder. To make a notation simpler, we define:
\begin{equation}
[\bv]_{p} \eqdef \sum_{i=1}^{p} b_i 2^{p-i},
\end{equation}for any binary vector $\bv=(b_1,...,b_p)$. Then, the user $k$'s channel input message at time slot $t$ is obtained as
\begin{equation}
w_k[t]  = [(\tau_{k}[pt],\tau_{k}[pt-1],...,\tau_{k}[pt-p+1])]_{p},
\end{equation}
for $t\in [n/p]$, where $n$ is assumed to be a multiple of $p$. As illustrated in Fig.~\ref{codedmodel}, each user $k$ transmits the message $\{w_k[t]: t \in [n/p]\}$ to the BS during the $n/p$ time slots. At each time slot $t \in [n/p]$, the BS computes the LLRs from the current observation $\rv[t]$ as
\begin{align}
L_{pt-(i-1)}^{k}(\rv[t]) &\eqdef \log\frac{\PP(\tau_k[pt-(i-1)]=0|\rv[t])}{\PP(\tau_k[pt-(i-1)]=1|\rv[t])}\nonumber\\
&= \log\frac{\sum_{\bv \in\{0,1\}^p:b_i = 0} \PP(w_{k}[t] = [\bv]_{p} | \rv[t] )}{\sum_{\bv\in\{0,1\}^p:b_i = 1} \PP(w_{k}[t] = [\bv]_{p} | \rv[t])},\nonumber
\end{align} for $i\in[p]$. In \cite{Hong-Soft}, it was shown that the above LLRs can be efficiently computed using the weighted Hamming distance in Definition 1 as
\begin{align}
L_{pt-(i-1)}^{k}(\rv[t]) &= \min_{ \cv_{\ell} \in \Bc_{(i,1)}^k} d_{{\rm wh}}\left(\rv[t],\cv_{\ell} ; \left\{\log\epsilon_{\ell,i}^{-1}\right\}\right)\nonumber\\
& -\min_{ \cv_{\ell} \in \Bc_{(i,0)}^k} d_{{\rm wh}}\left(\rv[t],\cv_{\ell} ; \left\{\log\epsilon_{\ell,i}^{-1}\right\}\right),\label{eq:LLRs}
\end{align}where $\epsilon_{\ell,i}$ is given in (\ref{eq:cross_p}) and the associated subcodes are defined as
\begin{equation}
\Bc_{(i,j)}^k=\bigcup_{\bv \in \{0,1\}^p :b_i = j} \Cc_{|\{w_k[t]= [\bv]_{p}\}} \mbox{ for } j\in\{0,1\}. \label{eq:1stcode}
\end{equation} During the $n/p$ time slots, the BS collects the LLRs $\{L_{pt-(i-1)}^{k}(\rv[t]): i\in[p], t\in[n/p]\}$ for $k=1,...,K$ and they are embedded into the corresponding channel decoder  $k$, for $k=1,...,K$.

%If the BS is equipped with one channel decoder, then all the $K$ users' messages are decoded sequentially.

%%%%%%%%%%%%%%%%%%%%%%%%%%%%%%%%
\section{The Proposed  SCSO Detector} \label{sic}

In this section, we present the SCSO detector which can enhance the SO detector in Section~\ref{sec:swmd}, by exploiting a {\em priori} knowledge conveyed by a channel decoder. The overall structure of the proposed detector is depicted in Fig.~\ref{codedmodel}. Furthermore, we develop an efficient greedy algorithm to optimize a decoding order. The corresponding detector is refereed to as an {\em ordered} SCSO detector.

%%%%%%%%%%%%%%%%%
\subsection{The SCSO Detector}

\begin{figure}[t]
\centerline{\includegraphics[width=9.5cm]{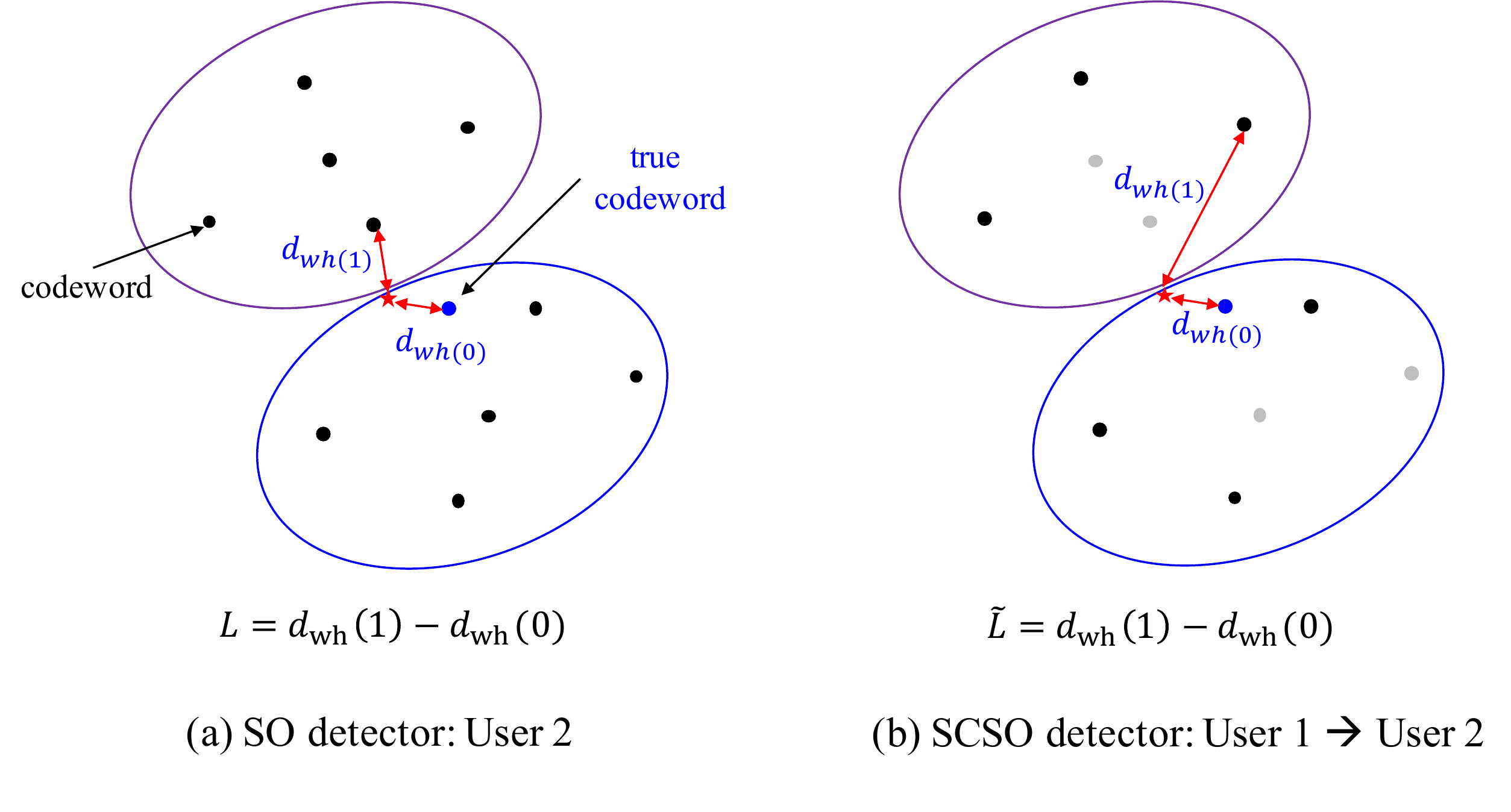}}\vspace{-0.5cm}
\caption{Illustration of LLR comparison between the SO detector and the SCSO detector. According to the previously decoded messages, some codewords (represented by grey circles) are eliminated from the code $\Cc$.}
\label{distance}
\end{figure}

Recall that in the SO detector, LLRs are computed by searching all the codewords in the $\Cc$ (see (\ref{eq:LLRs})). In contrast, the proposed SCSO detector produces the LLRs using a {\em refined} code, where the refined code (or subcode) contains some part of the codewords in the $\Cc$ according to the previously decoded messages, as shown in Fig.~\ref{distance}. Since in the subcode, the distances among the codewords are larger than those in the $\Cc$, an ambiguity of the detection can be reduced.

The detailed procedures of the SCSO detector are provided as follows. Focus on the LLR computations of the channel decoder $k$, with the knowledge of the $\left[k-1\right]$ users' decoded messages $\hat{\wv}_1^{k-1}[t]=\{\hat{w}_1 [t], \ldots, \hat{w}_{k-1} [t]\}$. We first define a {\em refined} subcode of the $\Cc$ as
\begin{align}
&\Cc_{\mid \{\wv_1^{k-1} [t]= \hat{\wv}_1^{k-1}\}} \nonumber \\
&\;\;\;\;\;\;\;\;\;\;\;\triangleq\{\cv=\Ec(\wv) : \wv \in \Wc, \wv_1^{k-1} [t]= \hat{\wv}_1^{k-1}[t]\}, \label{eq:codedef2}
\end{align} where its size is equal to $|\Cc|/2^{k-1}$. Since the distances of the remaining codewords can be larger as $k$ increases, the SCSO detector can produce more reliable LLRs as the cancellation step proceeds (see Fig.~\ref{distance}). Then, using the refined subcode
and  (\ref{eq:LLRs}), the {\em enhanced} LLRs are obtained as
\begin{align}
&\tilde{L}_{pt-(i-1)}^{k}(\rv[t],\;\hat{\wv}_1^{k-1}[t]) = \nonumber \\
&\;\;\;\;\;\;\;\;\;\;\;\;\min_{\cv_{\ell} \in \Bc_{\left(i,1\mid\hat{\wv}_1^{k-1}[t]\right)}^k} d_{{\rm wh}}\left(\rv[t],\cv_{\ell} ; \left\{\log\epsilon_{\ell,i}^{-1}\right\}\right)\nonumber\\
&\;\;\;\;\;\;\;\; -\min_{ \cv_{\ell} \in \Bc_{\left(i,0\mid\hat{\wv}_1^{k-1}[t]\right)}^k} d_{{\rm wh}}\left(\rv[t],\cv_{\ell} ; \left\{\log\epsilon_{\ell,i}^{-1}\right\}\right),\label{eq:LLRs2}
\end{align}
where the refined subcodes are constructed according to the known messages as
\begin{align}
&\Bc_{\left(i,j\mid\hat{\wv}_1^{k-1}[t]\right)}^k \nonumber \\
&\;\;\;\;\;\;\;\;\;\;=\bigcup_{\bv \in \{0,1\}^p :b_i = j} \Cc_{|\{w_k[t]= [\bv]_{p},\;\wv_1^{k-1} [t]= \hat{\wv}_1^{k-1}[t]\}}. \label{eq:reducedcode}
\end{align}
By embedding the above LLRs $\{\tilde{L}_{pt-1}^k(\rv[t],\;\hat{\wv}_1^{k-1}[t]): i\in[p], t\in[n/p]\}$ into the  channel decoder $k$, the user $k$'s message is decoded as the bit-stream  $\hat{\bv}_k$. Also, the the $\{\hat{w}_{k}[t]: t\in[n/p]\}$ are obtained from $\hat{\bv}_k$, using the channel encoder and modulation function. Leveraging the decoded messages $\hat{\wv}_1^{k-1}[t]$ and $\hat{w}_k[t]$, then, the enhanced LLRs for the channel decoder $k+1$ are computed. This process is repeatedly applied to (\ref{eq:codedef2})-(\ref{eq:reducedcode}) until all the $K$ users' messages are decoded. 

\begin{remark} The complexity of LLR computation is proportional to the size of associated subcodes because the detector need to search them for the minimum operation in (\ref{eq:LLRs}) and (\ref{eq:LLRs2}). Since the SO detector examines every codewords for each user, $K|\Cc|$ number of comparison is needed. However, the SCSO detector reduces the size of subcodes by half each time. Therefore, total number of comparison is $|\Cc|+|\Cc|/2+\cdots+|\Cc|/2^{K-1}\approx2|\Cc|$. The ratio of the SCSO detector to the SO detector in the total number of comparison is $2/K$, so the complexity decreases as $K$ increases.

\end{remark}

%%%%%%%%%%%%%%%%%%%
\subsection{The Ordered SCSO Detector}

We present an ordered SCSO detector which further improves the SCSO detector by carefully determining a decoding order. First of all, we notice that due to the non-linearity of the effective channel, it is not possible to use the SNR-based ordering in V-BLAST  \cite{Wolniansky}. In the proposed method, we determine a decoding order in a greedy fashion: for each decoding step $i$, a user index $k_i$ is chosen from the remaining user indices such that the distance between two disjoint subcodes, where one is associated with 
 $\tau_k[t]=0$ and the other is associated with $\tau_k[t]=1$, is maximized. This is motivated by the fact that LLRs with higher reliability tends to be obtained from far-off subcodes as if a lower-rate channel code performs better than a higher-rate channel code. In detail, a decoding order is determined as follows.

 A user index $k_1\in[K]$ to be decoded at the first step is chosen as

\begin{align}
&k_1 = \nonumber\\
 &\argmax_{k \in [K]} \sum_{i \in [p]} \left\{\mbox{D}_{\rm s}\Big(\bigcup_{[\bv]_p:b_i=0}\Cc_{\mid w_k=[\bv]_{p}},\bigcup_{[\bv]_p:b_i=1}\Cc_{\mid w_k=[\bv]_{p}} \Big)\right\}, \label{eq:1stselect}
\end{align} where $\mbox{D}_{s}(\Cc_1,\Cc_2)$ represents the set distance between two codes $\Cc_1$ and $\Cc_2$. It is not obvious to find an optimal set distance $\mbox{D}_{s}(\cdot,\cdot)$, which is left for a future work. For the time being, we resort to using the mean distance (e.g., distance between the centroids) as
\begin{equation}
\mbox{D}_{s}(\Cc_1,\Cc_2)\eqdef \left|\frac{1}{|\Cc_1|}\sum_{\cv \in \Cc_1}\cv - \frac{1}{|\Cc_2|}\sum_{\cv \in \Cc_2}\cv\right|^2.
\end{equation} 

Next, a user index $k_2\in[K]\setminus\{k_1\}$ to be decoded in the second step is chosen using the previously decoded user $k_1$'s message
 $\{\hat{w}_{k_1}[t]: t=1,...,n/p\}$. Since $\hat{w}_{k_1}[t]$ can have a different value for various $t$, the corresponding best user index can be different. Only one user index, however, should be selected for all time slots, in order to perform a sequential channel decoding. To meet this requirement, we select the user index to be chosen most frequently for the $n/p$ time slots. Namely, we choose a $k_2$ as
\begin{align*}
&k_2(\hat{w}_{k_1}) = \argmax_{k \in [K]\setminus\{k_1\}} \sum_{i \in [p]} \Big\{\mbox{D}_{\rm s}\Big(\bigcup_{[\bv]_p:b_i=0}\Cc_{\mid w_k=[\bv]_{p}, w_{k_1}=\hat{w}_{k_1}},\nonumber\\
 &\;\;\;\;\;\;\;\;\;\;\;\;\;\;\;\;\;\;\;\;\;\;\;\;\;\;\;\;\;\;\;\;\;\;\; \bigcup_{[\bv]_p:b_i=1}\Cc_{\mid w_k[t]=[\bv]_{p}, w_{k_1}=\hat{w}_{k_1}} \Big)\Big\},
\end{align*} where $\hat{w}_{k_1}=\mbox{Majority}\{\hat{w}_{k_1}[t]: t=1,...,n/p\}$ represents the most frequent values in 
$\{\hat{w}_{k_1}[t]: t=1,...,n/p\}$.

Likewise, a user index $k_{\ell}$ to be decoded at the $\ell$-th step is selected using the previously decoded messages $\{\hat{\wv}_{k_1}^{k_{\ell-1}}[t]: t=1,...,n/p\}$ as
\begin{align}
&k_{\ell}(\hat{\wv}_{k_1}^{k_{\ell-1}}) =\nonumber\\
& \argmax_{k \in [K]\setminus\{k_1,...,k_{\ell-1}\}} \sum_{i \in [p]} \Big\{\mbox{D}_{\rm s}\Big(\bigcup_{[\bv]_p:b_i=0}\Cc_{\mid w_k=[\bv]_{p}, \wv_{k_1}^{k_{\ell-1}}=\hat{\wv}_{k_1}^{k_{\ell-1}}},\nonumber\\
 &\;\;\;\;\;\;\;\;\;\;\;\;\;\;\;\;\;\;\;\;\;\;\;\;\;\;\;\;\;\;\;\;\;\;\; \bigcup_{[\bv]_p:b_i=1}\Cc_{\mid w_k[t]=[\bv]_{p},  \wv_{k_1}^{k_{\ell-1}}=\hat{\wv}_{k_1}^{k_{\ell-1}}} \Big)\Big\}, \label{eq:generalselect}
\end{align}where $\hat{w}_{k_i}=\mbox{Majority}\{\hat{w}_{k_i}[t]: t=1,...,n/p\}$ for $i=1,...,\ell-1$. 

The above process is repeatedly applied to all the $K$ users and the subsequent process is exactly identical with the proposed SCSO detector (see Algorithm 1).

\begin{algorithm}
\makeatletter
\newcommand{\algrule}[1][.2pt]{\par\vskip.5\baselineskip\hrule height #1\par\vskip.5\baselineskip}
\makeatother
    \caption{The ordered SCSO detector}
    \label{euclid}
    \begin{algorithmic}% The number tells where the line numbering should start
%        \Procedure{Euclid}{$a,b$} \Comment{The g.c.d. of a and b}
\State $\rhd$Choose $K$ and $N_r$ for MU-MIMO system 
\State $\rhd$Choose $p$ for $2^p(=m)$ QAM constellation
\State $\rhd$Choose blocklength $n$ and rate $R$ for channel coding
\algrule
\State Define the code $\Cc=\{\cv_{\ell}: \ell=0,...,m^K-1\}$ in (\ref{eq:code})
\State  One-bit quantized observation $\rv[t]$
            \For{$i = 1,\ldots, K$}            
                \If{$i=1$}  \Comment{No previous message}
                    \State Select $k_1$ from (\ref{eq:1stselect})
                    \State Compute LLRs using (\ref{eq:LLRs})-(\ref{eq:1stcode}) and the code $\Cc$
                \Else
                    \State Select $k_{i}$ from (\ref{eq:generalselect})
                    \State Compute LLRs using (\ref{eq:LLRs2})-(\ref{eq:reducedcode}) and the refined code 
                    \[\Cc_{\mid\{\wv_{k_1}^{k_{i-1}}=\hat{\wv}_{k_1}^{k_{i-1}}\}}\]
                \EndIf

            \State Decode the bit-stream $\bv_{k_i}[t]$ using the above LLRs
	    \State Re-encode the decoded bit-strem to get $\wv_{k_i}[t]$
            \State $\hat{w}_{k_{i}}=\mbox{Majority}\{\hat{w}_{k_{i}}[t]:t\in[n/p]\}$
            \State Define the refined code as  $\Cc_{\mid\{\wv_{k_1}^{k_{i}}=\hat{\wv}_{k_1}^{k_{i}}\}}$
            \EndFor
%            \State \textbf{return} $b$
%        \EndProcedure
    \end{algorithmic}
\end{algorithm}

%%%%%%%%%%%%%%%
\section{Numerical Results}

In this section, we compare the proposed ordered SCSO detector with the other detectors for the polar coded MU-MIMO systems with one-bit ADCs.  A Rayleigh fading channel is assumed in which each element of a channel matrix $\Hm$ is drawn from an independent and identically distributed (i.i.d.) circularly symmetric complex Gaussian random variable with zero mean and unit variance.  We adopt a polar code of the blocklength 128 (e.g., $n=128$) in \cite{Arikan}. The soft inputs (e.g., LLRs) of the polar decoder are computed from (\ref{eq:LLRs}) and (\ref{eq:LLRs2}) for SO and ordered SCSO detectors, respectively.  For the simulation, a rate $1/2$ polar code is used and SC polar decoder is used (see \cite{Arikan} for details). We would like to notice that similar trends in Figs.~\ref{R1} and~\ref{R2} are observed when using a more powerful SC list (SCL) polar decoder in \cite{Tal}. We compare the coded MU-MIMO systems with various parameters $K$ and $N_r$. Each simulation results shows the coded frame-error-rate (FER) performance where the FER represents the average FER over $K$ users. A perfect channel state information is assumed for all simulations.

 %For an polar code decoder,  bit-flipping decoder \cite{Rao} is used for ZF-type detectors in \cite{Choi} and belief-propagation decoder \cite{Richardson} is employed for the SO detector and the proposed SCSO detector. 

%Also, from (\ref{eq:LLRs}), the LLRs (i.e., soft inputs) of the belief-propagation decoder are computed  as 
%\begin{align*}
% L_{2t-1}^{k}(\rv[t])=\min_{\cv_{\ell} \in  \Cc_{|w_k[t]=2}\; \cup \; \Cc_{|w_k[t]=3}} d_{{\rm wh}}(\rv[t],\cv_{\ell}; \{\log\epsilon_{\ell,i}^{-1}\})\
%   - \min_{\cv_{\ell} \in  \Cc_{|w_k[t]=0}\;  \cup\;   \Cc_{|w_k[t]=1}} d_{{\rm wh}}(\rv[t],\cv_{\ell};\{\log\epsilon_{\ell,i}^{-1}\})\\
%L_{2t}^{k}(\rv[t])= \min_{\cv_{\ell} \in  \Cc_{|w_k[t]=1}\; \cup \; \Cc_{|w_k[t]=3}} d_{{\rm wh}}(\rv[t],\cv_{\ell} ;\{\log\epsilon_{\ell,i}^{-1}\})\\
% - \min_{\cv_{\ell} \in  \Cc_{|w_k[t]=0}\; \cup \; \Cc_{|w_k[t]=2}} d_{{\rm wh}}(\rv[t],\cv_{\ell} ;\{\log\epsilon_{\ell,i}^{-1}\}),
%\end{align*}for $t\in[n/2]$. 

%%%%%%%%%%%%%%%%%%%%%%%%
\subsection{Gain of successive cancellation}

Fig.~\ref{R1} shows the performance improvement of the proposed ordered SCSO detector over the SO detector in \cite{Hong-Soft} with respect to the number of receive antennas $N_r$. From this simulation, we observe that the proposed detector can provide a suitable coded gain over the SC detector. For a target FER (e.g., $10^{-1}$), we can see that the performance gap between the proposed detector and the SO detector becomes larger as $N_r$ decreases. When $N_r$ is smaller for a fixed $K$, the length of each codeword in the $\Cc$ is smaller and the number of $|\Cc|$ codewords are more densely located. Therefore, a refined subcode has more effect on the computation of LLRs.

%%%%%%%%%%%%%%%%%%%%%%%
\subsection{Comparison with other detectors}

\begin{figure}
\centerline{\includegraphics[width=9cm]{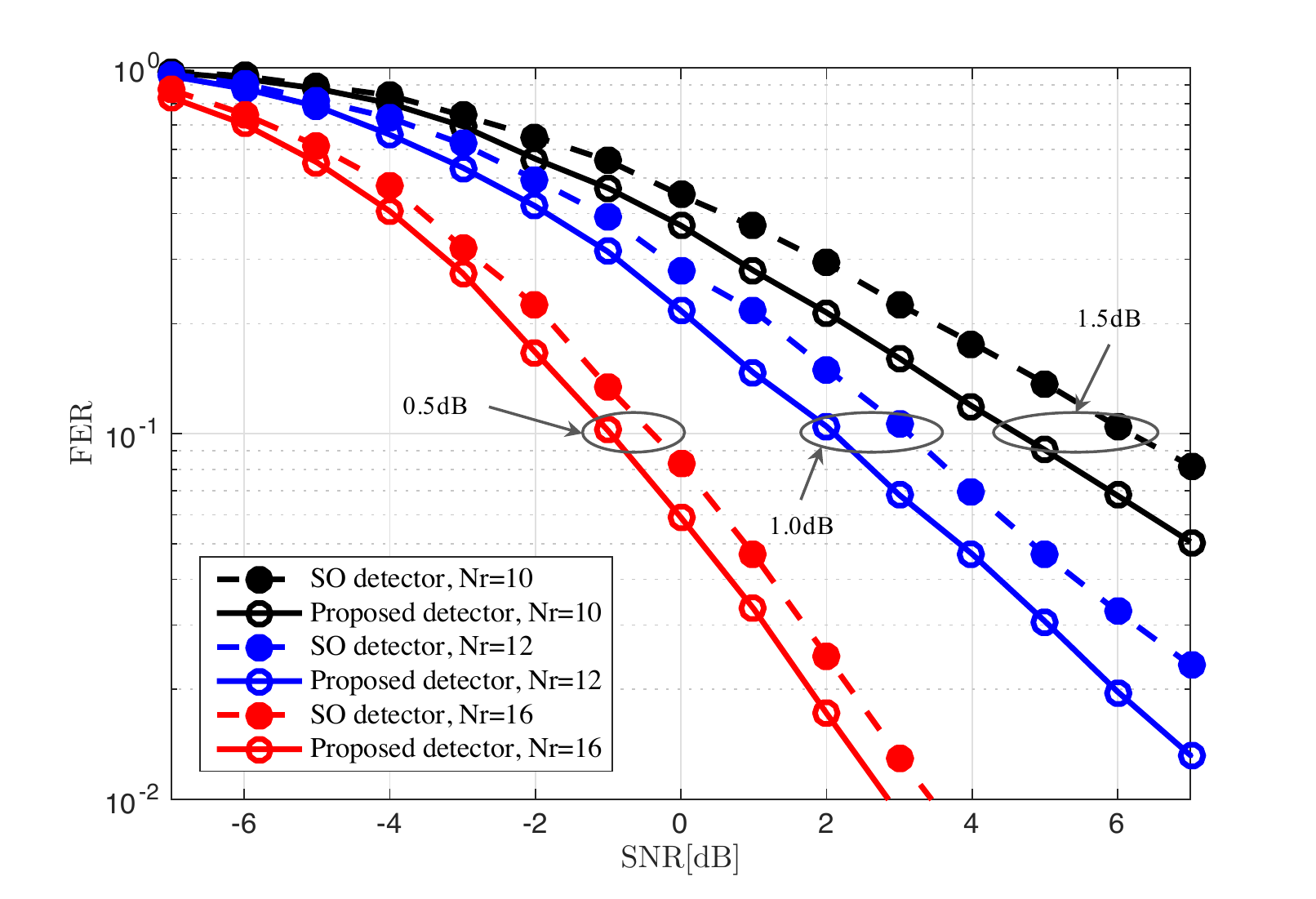}}\vspace{-0.5cm}
\caption{$K=6$. Performance comparisons of the SO and ordered SCSO detectors as a function of number of receive antennas $N_r$ for the polar-coded MU-MIMO system with one-bit ADCs.}
\label{R1}
\end{figure}

\begin{figure}
\centerline{\includegraphics[width=9cm]{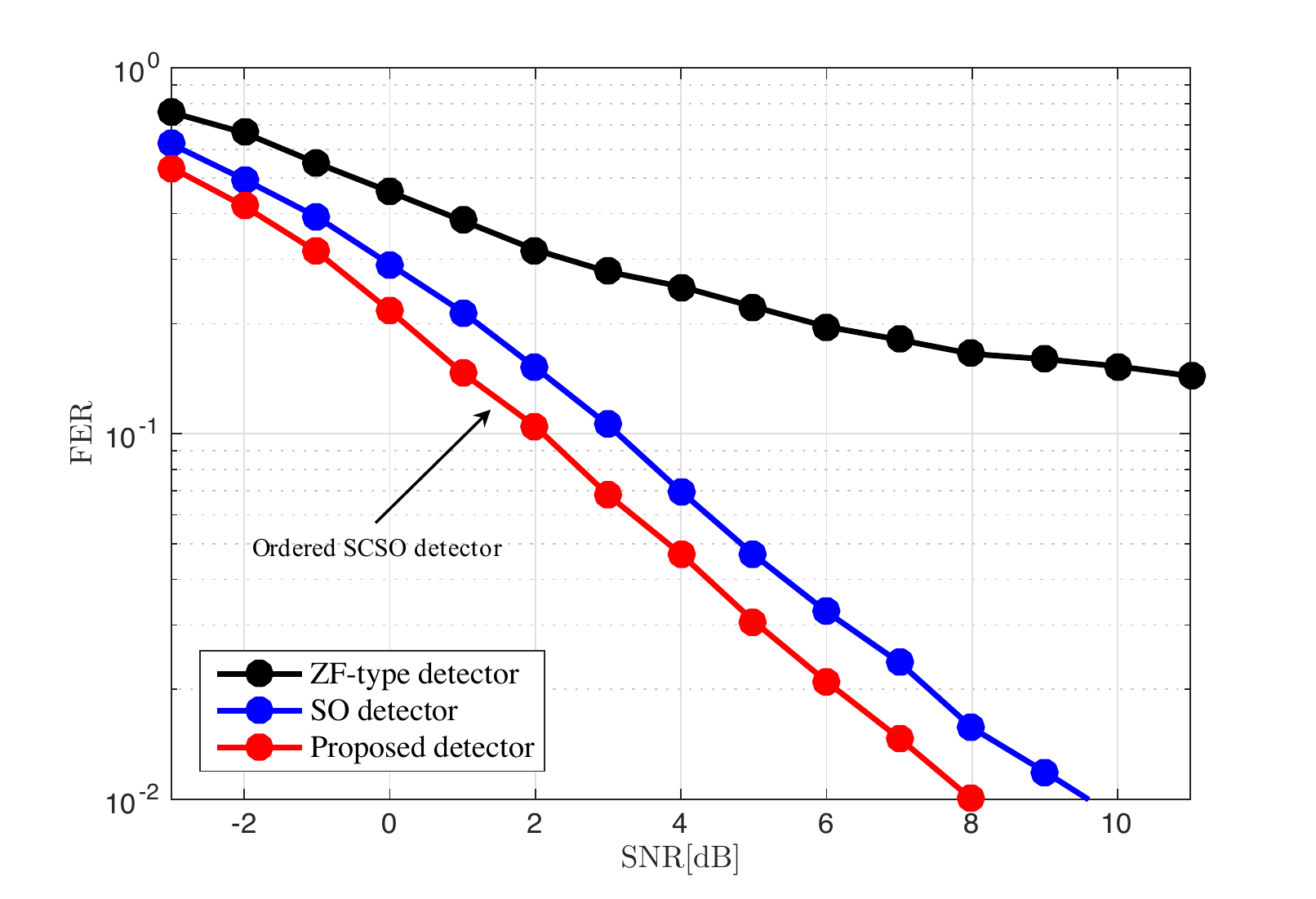}}\vspace{-0.5cm}
\caption{$K=6$ and $N_r=12$. FER comparisons of various {\em soft-output} detectors for MU-MIMO systems with one-bit ADCs.}
\label{R2}
\end{figure}

Fig.~\ref{R2} shows the coded FER performances for various soft-output MIMO detectors as the ZF-type detector \cite{Choi}, SO detector \cite{Hong-Soft}, and proposed ordered SCSO detector. Note that ML detector \cite{Choi}, near-ML detector \cite{Choi}, and supervised-learning detector \cite{Lee} are excluded in the comparison, since they cannot produce soft-outputs. Also, as already shown in \cite{Hong-Soft}, their performances are much worse than that of SO detector. We observe that the proposed detector significantly outperforms the ZF-type detector and the gap becomes larger as SNR increases. As observed in Figs.~\ref{R1} and~\ref{R2} it provides the additional coded gain over the SO detector. As noticed in Remark 2, the complexity of the proposed detector can be reduced as similar to that of ZF-type detector, by maintaining the performance. Due to its good performance and low-complexity, therefore, the proposed detector can be a good candidate as a MIMO detector for the uplink MU-MIMO system with one-bit ADCs.

%%%%%%%%%%%%%%%%%%%%%%%%%%%%
\begin{remark} The computational complexities of the both SO and SCSO detectors are problematic for a large $K$ since the size of the code $\Cc$ (i.e., search-space) grows exponentially with $K$. Recently, the authors in \cite{SH} and \cite{Jeon} developed the efficient methods to significantly reduce the search-space, where the reduced search-space only takes the codewords from the $\Cc$ that lie inside the sphere centered at the current observation $\rv[t]$ with a certain radius. Since these methods can be directly applied to the both SO and SCSO detectors, their complexities can be considerably reduced.
\end{remark}

%%%%%%%%%%%%%%%%%%%%%%%%%%%%%%%%%%%%%%%%%%%%%%%%%

%Fig.~\ref{compareOrder2} shows the comparison of the latency in the SIC detector and soft-output detector. We measure time consumption in detecting $T$ number of messages for each user. Such result demonstrates that the proposed SIC detecor has even less time consumption. To detect $T$ messages, the soft-output detector must search $T|\Cc|$ subcodes. However, the CS detector searches at last $\frac{T|\Cc|}{2\{1-\left(\frac{1}{2}\right)^K\}K}$ subcodes because size of possible subcodes continues to shrink in half within $K$ users. This trend is enhanced as $K$ increases.

%%%%%%%%%%%%%%%%%%%%%%%%%%%%%%%%%%%%%%%%%%%%%%%%%%

\section{Conclusion}\label{sec:con}

We proposed the ordered successive-cancellation-soft-output (SCSO) detector for MU-MIMO systems with one-bit ADCs. The main idea of the proposed detector is that the previously decoded messages (fed by the channel decoders) are exploited to improve the reliabilities of the soft inputs of a subsequent channel decoder. Furthermore, we developed an efficient greedy algorithm to find a good decoding order. Via simulation results, it was shown that the ordered SCSO detector significantly outperforms the other detectors for the coded MU-MIMO systems with one-bit ADCs. One possible extension of this work is to optimize a distance measure between two subcodes defined over Hamming space.

%%%%%%%%%%%%%%%%%%%%%%%%%%%%%%%%%%%%%%%%%%%
%%%%%%%%%%%%%%%%%%%%%%%%%%%%%%%%%%%%%%%%%%%

%%%%%%%%%%%%%%%%%%%%%%%

%\bibitem{Hagenauer} J. Hagenauer, ``The turbo principle: Tutorial introduction and state of the art," in {em Proc. 1st Int. Symp.  Turbo Codes,} pp. 1-12, Sept. 1997.

%%%%%%%%%%%%%%%%%%%%%%%

%%%%%%%%%%%%%%%%%%%%%%%%%%%%%%%%%%%%%%%%%%%%
\end{document}